\documentstyle[preprint,aps]{revtex}
\begin{document}
\draft
\preprint{UCI-TR 2001-19}
\title{Gravity in Dynamically Generated Dimensions}
\author{Myron Bander\footnote{Electronic address: mbander@uci.edu}
}
\address{
Department of Physics and Astronomy, University of California, Irvine,
California 92697-4575}

\date{July\ \ \ 2001}
\maketitle
\begin{abstract}
A theory of gravity in $d+1$ dimensions is dynamically generated from a
theory in $d$ dimensions. As an application we show how $N$ dynamically
coupled gravity theories can reduce the effective Planck mass.
\end{abstract}


\section{Introduction}\label{intro}
The idea that a gauge theory in $(1,d)$ dimensions appears as the low energy
limit of a $(1,d-1)$ theory with many fields has been recently put forward
\cite{a-hcg,hpw}. One starts with $(1,d-1)$ gauge fields $A_\mu(x,i)$,
$\mu=0,1,\cdots, d-1$ with the range of the discrete index $i$ being either
infinite of finite and periodic ($A_\mu(x,1)=A_\mu(x,N)$). Interactions are
chosen as to insure that a field $A_{i,i+1}(x)$ is generated dynamically
and whose interactions, in the low energy limit, mimic $A_d$ in a $(1,d)$
dimensional space with $i$ turning into the discrete extra dimension.  A
feature of this approach is that the $(1,d-1)$ theory has the desired
properties of renormalizability and asymptotic freedom.

In this work, we extend this approach to gravity, namely we generate a
$(1,d)$ dimensional gravity from a theory in $(1,d-1)$ dimensions. Gravity
will be described in a moving frame formalism as an $SO(1,d-1)$ gauge theory
on a $(1,d-1)$ manifold.  Unlike the situations discussed in
\cite{a-hcg,hpw}, where one started with a gauge theory in some dimension
and generated the same gauge theory in a space with one higher dimension,
in order to generate gravity in the higher dimension we have to start with
an extended gravity in the lower dimensional space. Namely, we start with an
$SO(1,d)$ gauge theory on a $(1,d-1)$ manifold and dynamically generate the
$SO(1,d)$ theory on a $(1,d)$ manifold \cite{as}. Of course, the lower
dimensional theory includes gravity in the $SO(1,d-1)$ subgroup of
$SO(1,d)$.

In this case we cannot appeal to renormalizability or to asymptotic freedom
to justify this approach as the lower dimensional gravity or extended
gravity is unlikely to be renormalizable or asymptotically free. What our
construction insures is that the dynamically generated higher dimensional
theory is no more singular that the lower dimensional one and that
coordinate invariance and local $SO(1,d)$ invariance are maintained at each
step. Details, as well as a discussion of $SO(1,d)$ invariant interactions
in $(1,d-1)$ dimensional spaces, are presented in Sec.\ \ref{details}. In
Sec.\ \ref{ED} we apply this approach to a scenario where in four
dimensions the existence of $N$ dynamically coupled gravity theories
decreases the effective gravitational coupling by a factor of $N$.

\section{Dynamical Generation of Gravity}\label{details}
\subsection{Gravity in $d+1$ Dimensions}
We shall first discuss gravity theory on a $(1,d)$ dimensional manifold
with coordinates $x_\mu;\ \mu=0,1,\cdots,d$, which we wish to obtain through
the dynamical generation of a dimension in some theory on a $(1,d-1)$
dimensional manifold. Our goal is the $(1,d)$ Einstein-Hilbert Lagrangian,
which we express in the moving frame or d-ad formalism.
\begin{eqnarray}
L_{EH}=&M&_{(d+1)}^{d-1}
 \epsilon^{\mu_1\mu_2\cdots\mu_{d+1}}\epsilon_{a_1a_2\cdots a_{d+1}}
   \left[R_{\mu_1\mu_2}^{a_1a_2}(x)+M^2_{(d+1)}\lambda_{(d+1)}
 e_{\mu_1}^{a_1}(x)e_{\mu_2}^{a_2}(x)\right]\nonumber \\
   &\times&e_{\mu_3}^{a_3}(x)e_{\mu_4}^{a_4}(x)\cdots 
    e_{\mu_{d+1}}^{a_{d+1}}(x)\, ;\label{ehlagrangian}
\end{eqnarray}
in the above the $e_\mu^a(x)$'s, with flat space, Minkowski, indexes
$a=0,1,\cdots,d$, are the (d+1)-ads and the $\omega_\mu^{ab}(x)$'s are the
spin-connections. $M_d$ is the $d$ dimensional Planck mass, $M_d^2\lambda_d$
is a  cosmological constant and the curvature tensor 
$R_{\mu_1\mu_2}^{a_1a_2}(x)$ is related to the spin-connections by
\begin{equation}\label{curvature}
R_{\mu\nu}^{ab}(x)=\partial_\mu\omega_\nu^{ab}(x)
  -\omega_{\mu,n}^{a}(x)\omega_\nu^{nb}(x)-(\mu \leftrightarrow \nu)\, .
\end{equation} 
In order to see what theory in a $(1,d-1)$ space we should start with, 
we foliate the $(1,d)$ dimensional manifold into $(1,d-1)$ dimensional
ones. Specifically we single out the last coordinate $x_d$. Coordinates in
the $(1,d)$ dimensional space are written as $(x_\mu,x_d)$, where now
$\mu=0,1,\cdots, d-1$. We leave the first $d$ d-ads as they were but
separate out the ``shift'' vector \cite{mwt}
\begin{equation}\label{mwt}
e_d^a(x)=N^a(x)\, ;
\end{equation}
the Minkowski index, $a$, still ranges over (d+1) values. In terms of this
shift vector, eq.(\ref{ehlagrangian}) becomes \cite{mb}
\begin{equation}
L_{EH}=L_A+L_B\, ,
\end{equation}
with
\begin{eqnarray}
L_A&=&M_{(d+1)}^{d-1}
 \epsilon^{\mu_1\mu_2\cdots\mu_{d}}\epsilon_{a_1a_2\cdots a_{d+1}}
   \left[(d-1)R_{\mu_1\mu_2}^{a_1a_2}(x)+
     (d+1)M^2_{(d+1)}\lambda_{(d+1)}e_{\mu_1}^{a_1}(x)e_{\mu_3}^{a_3}(x)
      \right] \nonumber\\ \label{LA}
  &{}&{\times}e_{\mu_3}^{a_3}(x)e_{\mu_4}^{a_4}(x)
      \cdots e_{\mu_{d}}^{a_{d}}(x)N^{a_{d+1}}(x)\, ,\\ 
L_B&=&2M_{(d+1)}^{d-1}\label{LB}
  \epsilon^{\mu_1\mu_2\cdots\mu_{d}}\epsilon_{a_1a_2\cdots a_{d+1}}
   R_{d\mu_1}^{a_1a_2}(x)
     e_{\mu_2}^{a_3}(x)e_{\mu_3}^{a_4}(x)\cdots e_{\mu_{d}}^{a_{d+1}}(x)
      \, .
\end{eqnarray}
We note that $L_A$ does not contain any derivatives in the $x_d$ direction
nor any terms involving $\omega_{d\mu}^{ab}(x)$ while $L_B$ does but, in
turn, does not involve the shift vectors $N^a(x)$. {\em $L_A$ will
determine the $(1,d-1)$ theory we start out with, and our goal will be to
generate dynamically $L_B$.}

\subsection{$SO(1,d)$ Gauge Theory in $d$ Dimension}\label{(d+1)}
The Lagrangian $L_A$ in eq.\ (\ref{LA}) describes an $SO(1,d)$ gauge theory
in on a $(1, d-1)$ manifold; as $SO(1,d-1)$ is a subgroup of $SO(1,d)$, this
model includes gravity, an $SO(1,d-1)$ gauge theory on a $(1,d-1)$
manifold. We shall now obtain some of the properties of this extended
model; these properties are needed by other fields and their
interactions. Following eq. (\ref{LA}) we rewrite the Lagrangian as
\begin{eqnarray}
L_{SO(1,d)}=&M&_d^{d-2}
 \epsilon^{\mu_1\mu_2\cdots\mu_{d}}\epsilon_{a_1a_2\cdots a_{d+1}}
   \left[R_{\mu_1\mu_2}^{a_1a_2}(x)+M^2_d\lambda_d
   e_{\mu_1}^{a_1}(x)e_{\mu_2}^{a_2}(x)\right]\nonumber \\
     &\times&e_{\mu_3}^{a_3}(x)e_{\mu_4}^{a_4}(x)
       \cdots e_{\mu_d}^{a_d}(x)N^{a_{d+1}}(x)\, .\label{so(1,d)lagrangian}
\end{eqnarray}
As $e_\mu^a(x)$ is not a square matrix we may not define an $e^\mu_a(x)$ as
its inverse. We can, however, introduce a $d\times d$ metric tensor
\begin{equation}
g_{\mu\nu}(x)=e_{\mu,a}(x)e_\nu^a(x)
\end{equation}
as well as its inverse $g^{\mu\nu}(x)$, thus allowing us to raise and lower
the curved space coordinate indexes. Using the $SO(1,d)$ Clifford algebra
\begin{equation}
\{\gamma_a,\gamma_b\}=2\eta_{ab}
\end{equation}
and its associated spin matrices $\Sigma_{ab}=[\gamma_a,\gamma_b]/2i$ the
Dirac Lagrangian for an $SO(1,d)$ spinor field $\psi(x)$ is 
\begin{equation}
L_D= \epsilon^{\mu_1\mu_2\cdots\mu_{d}}\epsilon_{a_1a_2\cdots a_{d+1}}
       {\bar\psi}(x)\gamma^{a_1}D_{\mu_1}\psi(x)
  e_{\mu_2}^{a_2}(x)e_{\mu_3}^{a_3}(x)\cdots
e_{\mu_d}^{a_d}(x)N^{a_{d+1}}(x)
  \, ,
\end{equation}
with the covariant derivative 
\begin{equation}
D_\mu\psi(x)=\left(\partial_\mu-\frac{1}{2}\Sigma_{ab}\omega_\mu^{ab}\right)
    \psi(x)\, .
\end{equation}

\subsection{Dynamical Generation of Gravity in $d+1$ Dimensions}
In order to generate an extra dimension we study many mutually
non-interacting $SO(1,d)$ theories described by d-ads $e_\mu^a(x,i)$,
spin connections $\omega_\mu^{ab}(x,i)$ and fields $N^a(x,i)$; at this
point the range of the $i$'s need not be specified. The Lagrangian for this
collection of theories is
\begin{eqnarray}\label{basiclagrangian}
L_0=&M&_d^{d-2}\sum_i
 \epsilon^{\mu_1\mu_2\cdots\mu_{d}}\epsilon_{a_1a_2\cdots a_{d+1}}
   \left[R_{\mu_1\mu_2}^{a_1a_2}(x,i)+M_d^2\lambda_d
     e_{\mu_1}^{a_1}(x,i)e_{\mu_2}^{a_2}(x,i)\right]\nonumber \\
 &\times&e_{\mu_3}^{a_3}(x,i)e_{\mu_4}^{a_4}(x,i)\cdots e_{\mu_d}^{a_d}(x,i)
     N^{a_{d+1}}(x,i)\, .
\end{eqnarray}
It is invariant under the product group $\cdots\times SO^i(1,d)\times
SO^{(i+1)}(1,d)\times\cdots$.

In order to couple theories at different $i$'s we have to introduce several
more fields. For each pair $(i,i+1)$ there is a nonabelian gauge field
$A_\mu^{i,i+1}(x)$; the only requirement on the group $G$ under which these
fields transforms and the strength of the gauge coupling is that certain
fermion condensates, to be discussed below, are induced. In addition, for
each $i$, we have two Weyl fermion fields. One, $\psi^i(x)$, couples to
$A_\mu^{i,i+1}(x)$ as a fundamental under $G$ while the other one,
$\chi^i(x)$, couples as an anti-fundamental under to $A_\mu^{i-1,i}(x)$. We
assume that the $SO^i(1,d)\times SO^{(i+1)}(1,d)$ symmetry is broken by a
condensate
\begin{equation}\label{condensate}
\langle \psi^i(x)\chi^{i+1}(x) \rangle\sim f_G^{d-1} 
   \exp \left[\frac{i}{2}\Sigma_{ab} {\cal O}_{i,i+1}^{ab}(x)\right]\, ;
\end{equation}
${\cal O}_{i,i+1}^{ab}(x)$ is an $SO(1,d)$ Lorentz transformation matrix
and $f_G$ parameterizes the strength of the condensate. The low energy
effective 
theory for the fields ${\cal O}_{i,i+1}^{ab}(x)$ is governed by the Lagrangian
\begin{equation}\label{generatedlagrangian}
L_1=2f_G^{d-1}\sum_i\epsilon^{\mu_1\mu_2\cdots\mu_{d}}\epsilon_{a_1a_2\cdots
   a_{d+1}} 
   {\cal O}_{i,i+1}^{a_1n}(x)D_{\mu_1} {\cal O}_{i,i+1;n}^{a_2}(x)
     e_{\mu_2}^{a_3}(x,i)e_{\mu_3}^{a_4}(x,i)\cdots
      e_{\mu_d}^{a_{d+1}}(x,i)\, ;
\end{equation}
the covariant derivative is 
\begin{equation}
D_\mu{\cal O}_{i,i+1}^{ab}(x)=\left[\partial_\mu{\cal O}_{i,i+1}^{ab}(x)
  +\omega^a_{\mu;n}(x,i){\cal O}_{i,i+1}^{nb}(x)-
    {\cal O}_{i,i+1}^{an}(x){\omega_{\mu;n}}^b(x;i+1)\right]\, .
\end{equation}
In the continuum limit we may expand ${\cal O}_{i,i+1}^{ab}(x)$
\begin{equation}
{\cal O}_{i,i+1}^{ab}(x)=\eta^{ab}+\frac{1}{a}\omega_{i,i+1}^{ab}(x)+
    \cdots \, ,
\end{equation}
where $a$ is the lattice separation. With the following identifications we
recover the discrete version of $L_B$ (eq.\ (\ref{LB})):
\begin{eqnarray}
a&=&{M_d^{d-2}}/{f_G^{d-1}}\, ,\nonumber \\
f_G&=&(d-1)M_{d+1}\, ,\\
M_d^2\lambda_d&=&(d+1)M_{d+1}\lambda_{d+1}\, .\nonumber
\end{eqnarray}
\section{Extra Discrete Dimensions}\label{ED}
Recently, extensive research has been carried out on the possibility that
extra compact but large dimensions may account for the apparently large
value of the Planck mass \cite{a-hdd,rs}. The present work shows how to
formulate a discrete version of such schemes. In the continuum case
phenomenology demands that we have more than one extra dimension. For
simplicity we shall discuss only one extra ``large'' discrete dimension
\cite{moredim}. We envisage a four dimensional manifold with many $SO(1,4)$
theories. The Lagrangian is the the sum of eq. (\ref{basiclagrangian}) and
(\ref{generatedlagrangian}) with $d=3$, $i=0,1,\ldots , N-1$, and periodic
conditions on the discrete index $i$,
$e_\mu^a(x,N)=e_\mu^a(x,0),\ldots$. All other non-gravity fields appear
only once and couple only to $e_\mu^a(x,0)$; in the continuum language this
would indicate that these extra fields do not propagate into the extra
dimension. The dynamical mechanism discussed in Sec.\ (\ref{details})
generates a fifth dimension of circumference $Na$. Using techniques similar
to those discussed in ref. \cite{a-hcg} we find that the potential for two
masses coupled only to the $i=0$ gravity is
\begin{equation}
V(r)=\frac{m_1m_2}{NM_4^2}\frac{1}{r}\sum_{m=0}^{N-1}\exp(-2\frac{r}{a}
\sin{\frac{m\pi}{N}})\, .
\end{equation}
For $r>>Na$ we recover the $1/r$ potential with an effective Planck
mass $M_P^2=NM_4^2$.

\end{document}